\documentclass[a4paper]{amsart}

\usepackage{graphics,amssymb,enumerate}
\usepackage{hyperref}
\usepackage[dvips]{graphicx}
\usepackage[all]{xy}

\def\r{ {\bf R}}

\newcommand {\bd} {\begin{displaymath}}
\newcommand {\ed} {\end{displaymath}}
\newcommand {\be} {\begin{equation}}
\newcommand {\ee} {\end{equation}}
\newcommand {\bea} {\begin{eqnarray}}
\newcommand {\eea} {\end{eqnarray}}

\def\pb#1{\left\{#1\right\}}
\def\PB{\left\{\cdot\,,\cdot\right\}}

\newtheorem{theorem}{Theorem}
\newtheorem{proposition}{Proposition}

\newtheorem{rem}{Remark}
\newtheorem{example}{Example}

\theoremstyle{definition}

\newcommand{\p}{\partial}

\Large
\begin{document}

\title{Poisson brackets after Jacobi and Pl\"ucker}

\author{ Pantelis A.~Damianou  }
\address{Department of Mathematics and Statistics\\
University of Cyprus\\
P.O.~Box 20537, 1678 Nicosia\\Cyprus}
\email{ damianou@ucy.ac.cy}

\date{}

\begin{abstract}
We construct a symplectic realization and a bi-hamiltonian formulation of a 3-dimensional system whose solution are the Jacobi elliptic functions.  We generalize this system and the related
Poisson brackets to higher dimensions.   These more general systems are parametrized by lines in  projective  space. For these rank 2 Poisson brackets  the Jacobi identity is satisfied only when the  Pl\" ucker relations hold.
Two of these Poisson brackets are compatible if and only if the corresponding lines in projective space intersect. We present several examples of such systems.
\end{abstract}

\maketitle

\Large
\section{Introduction}
\label{intro}

The main focus of this paper is the study of Poisson structures of the form

\be \label{poi}  \{ x_i,  x_j \} = \pi_{ij} x_1 x_2 \cdots \hat{x_i} \cdots \hat{x_j}  \cdots x_n   \ .  \ee

These  brackets are generalizations of the Sklyanin bracket which was defined in \cite{sklyanin} in dimension four.
As usual the notation $\hat{x_i}$ means that the variable $x_i$ is omitted.  We prove several properties of these  brackets:

\smallskip
\noindent
{\it i)} These brackets  are parametrized by lines in  projective  space.

\smallskip
\noindent
{\it ii)} The  bracket is Poisson  if and only if the functions $\pi_{ij}$ satisfy the Pl\" ucker relations.

\smallskip
\noindent
{\it iii)} The bracket is  decomposable, i.e., of the form   $\pi = \Psi  X \wedge Y$ where  $X$ and $Y$ are  vector fields and $\Psi=x_1 x_2 \dots x_n$.

\smallskip
\noindent
{\it iv)} The  rank of the Poisson structure  is  two.

\smallskip
\noindent
{\it v)} The functions
$f_{ijk}=\pi_{jk} x_i^2 -\pi_{ik} x_j^2 +\pi_{ij} x_k^2$ are Casimirs.

\smallskip
\noindent
{\it vi)} The bracket is equal (up to a constant multiple) to a  Jacobian type formula (Nambu-Poisson bracket) generated by $n-2$ Casimirs.

\smallskip
\noindent
{\it vii)} Two of these Poisson brackets are compatible if and  only if the corresponding lines in projective space intersect.

The simplest such example  occurs in dimension 3 and has the form

\bd \{x,y \}= \alpha z  \qquad \{ x, z \} =\beta y  \qquad \{ y, z \}=  \gamma  x  \ . \ed

Choosing a quadratic  Hamiltonian and suitable values of the parameters $\alpha, \beta, \gamma$  we obtain a completely integrable Hamiltonian system.
The   system of differential equations  is the following:

\begin{eqnarray} \label{e0}
\dot x &=& y z \\ \nonumber \dot y&=& -x z \\  \nonumber \dot z
&=& -k^2 x y  \ .
\end{eqnarray}

The solutions of the system turn out to be  the Jacobi elliptic functions
$sn(t,k)$, $cn(t,k)$, $dn(t,k)$.

We define a symplectic realization of the system from   ${\bf R}^4 $ to
$ {\bf R}^3$. The standard symplectic bracket in ${\bf R}^4 $ maps onto the three dimensional Poisson bracket.
The Hamiltonian in four dimensions  is of the form
\bd  H={1 \over 2} (1+k^2) p_1^2 +{ 1 \over 2} p_2^2 +{ 1\over 2}
\left( p_1 q_2 - p_2 q_1 \right)^2 \ .  \ed
This Hamiltonian generalizes naturally to ${\bf R}^6 $
and gives   a symplectic realization of the well-known  Clebsch system. The  Clebsch system describes the movement of a solid in an ideal fluid.

The Poisson bracket (\ref{poi})  can be obtained in a different way  using a Jacobian type Poisson formula. It is also called a Nambu-Poisson structure in \cite{camille}.
This type of  Poisson structure first appeared  in \cite{damianou89} in 1989  and was attributed  to H.
Flaschka and T. Ratiu. For explicit proofs see \cite{camille,  grabowski,  takhtajan}.   This formula may be obtained also from
Nambu brackets by fixing  some variables to constant.  Nambu defined this generalization of Poisson bracket in  three dimensional space $\r^3$ in 1973.
 The Nambu
brackets were generalized and given a geometrical interpretation by Takhtajan in \cite{takhtajan}  (see also,  \cite{damianou96,  damianou07}).
  Let us  point out how such a Jacobian Poisson structure is
defined. Let $f_1,\dots,f_{n-2}$ be $n-2$ independent functions in $n$  variables
$x_1,\dots,x_{n}$.  Then a  Poisson structure is defined on $\r^{n}$ by
\begin{equation}\label{det}
  \pb{f,g}_{det}:=\Psi \det(\nabla f,\ \nabla g,\ \nabla f_1,\ \dots,\ \nabla f_{n-2}),
\end{equation}%
where $\Psi$ is an  arbitrary smooth function. It is clear that each of the $f_i$ is a Casimir of $\PB_{det}$, so
that in particular the generic rank of $\PB_{det}$ is two. Note that if $\pi$ is a Poisson structure of rank two  then for every arbitrary  smooth function $\Psi$ the tensor $\Psi \pi$ is still a Poisson structure (see, \cite{camille}).

The system (\ref{e0}) is bi-hamiltonian, a fact which can be explained in the setting of Nambu mechanics. In his 1973 paper \cite{nambu}, Nambu introduced a generalization of Poisson bracket based on triples $\{f, g, h \}$ instead of $\{f, g \}$.    The evolution of systems is defined by  two Hamiltonians $H,G$  in the form
\bd \dot{f}=\{f, H, G \} \ . \ed
One obtains two Poisson brackets either by fixing $H$  (i.e. $\{f, g\}= \{f, g, H \}$ ) or by fixing $G$.  When we  fix $H$ the function $G$ plays the role of Hamiltonian and $H$ is a Casimir by skew-symmetry. When we fix $G$ the roles are reversed. The bi-hamiltonian formulation of Section 3 can be put in this setting.

 Background on elliptic functions is given in Section 2.  A symplectic realization and bi-hamiltonian formulation  of the system (\ref{e0}) is given in Section 3. Background material on Pl\"  ucker relations are presented in Section 4. In Section 5 we present in detail the example of system (\ref{poi})  in dimension 4. Finally, Section 6 studies system (\ref{poi}) in full generality.

\section{Jacobi Elliptic functions}
\label{jacobi}
Jacobi's elliptic functions have several applications in Geometry,
Mechanics and Physics. They are called elliptic since they appear
in some integrals needed to calculate the perimeter of the
ellipse. The formula for the length of an ellipse has the form

\bd \int_0^{ {\pi \over 2}} \sqrt {1 -k^2 \sin^2 \theta} \  d
\theta \  \ed
and the integral is usually called an  elliptic integral of the
second type.  Elliptic integrals of the first kind have the form

\bd F(t,k)=\int_0^{t} { dx \over \sqrt{ (1 -x^2) (1-k^2 x^2) }} =
\int_0^{ {\pi \over 2}} { d\theta \over \sqrt{ 1 -k^2 \sin^2 \theta }} \ ,  \ed

where $k \in (0,1)$.
Ever since the 1790's  Gauss observed that the inverse function of

\bd sin^{-1} (t)=\int_0^t  { dx \over \sqrt{ 1 -x^2} }  \ed
gives a  simply periodic function while the inverse function of
$F(t,k)$ is a doubly periodic function.

Thus, in the same manner that we define the sine as the inverse
function of this integral, we define the function
 $sn(k,t)$ as the inverse of  $F(t,k)$.
The elliptic functions which were discovered again later by Abel
and Jacobi are denoted by $sn(t,k)$, $cn(t,k)$ and $dn(t,k)$. The
real number $k$ takes values in the interval between $0$ and $1$.
These three functions satisfy the identities

\bd sn^2 (t,k)+ cn^2 (t,k)=1 \ed

\bd k^2 sn^2 (t,k) + dn^2 (t,k) =1 \ . \ed
These functions are generalizations of the trigonometric
functions in the sense that  $sn(t,0)=\sin t$ and  $cn(t,0)=\cos t$.
There are many ways to define trigonometric functions.  One
approach is to define  $\sin t$ and $\cos t$ as the solutions of the
system

\begin{eqnarray*}
{ dx \over dt}&=& y \\
{dy \over dt} &=& -x \ ,
\end{eqnarray*}

with initial conditions  $x(0)=0$, $x^{\prime} (0)=1$, $y(0)=1$,
$y^{\prime } (0)=0$.

The function $f(x,y)=x^2+y^2$ is a constant of motion of the above system, since \bd
\dot f=2 x \dot x + 2 y \dot y =2xy+2y (-x) =0 \ . \ed Therefore
we have  $x^2(t)+y^2(t)=c$.   Letting   $t=0$, we   conclude that $c=1$.
Therefore, since $x(t)=\sin t$ and $y(t)=\cos t$, we obtain  \bd \sin^2 t +\cos^2 t =1 \ . \ed

  Inspired by the previous remarks, and
following \cite{cushman, Meyer},  we define the functions
$sn(t,k)$, $cn(t,k)$, $dn(t,k)$ to be the solutions of the
equations

\begin{eqnarray} \label{e3}
\dot x &=& y z \\ \nonumber \dot y&=& -x z \\  \nonumber \dot z
&=& -k^2 x y  \ ,
\end{eqnarray}
with initial conditions $sn(0,k)=x(0)=0$, $cn(0,k)=y(0,k)=1$,
$dn(0,k)=z(0)=1$.  Right away we obtain the formulas for the derivatives of
these functions, e.g.

\bd {d \over dt} sn(t,k)= cn(t,k)\  dn(t,k) \ . \ed
It is easy to see that

\begin{eqnarray*}
sn(t,0)&=&\sin t \\
cn(t,0)&=&\cos t \\
dn(t,0)&=& 1 \ .
\end{eqnarray*}
  When  $k=0$,  then $z$ is constant and using the initial
conditions $z=1$. We end-up with the system $\dot x=y$, $\dot
y=-x$ which, as we know has the solution $x=\sin t$, $y=\cos t$.

   Equations   (\ref{e3}) have two constants of motion:

\bd F=x^2+y^2,  \qquad G=k^2 x^2 +z^2 \  .  \ed
As a corollary we obtain: For   $0 <k<1$, and for every real   $t$

\bd sn(t,k)^2 +cn(t,k)^2 =1 ,  \qquad k^2 sn(t,k)^2 + dn(t,k)^2=1  \  . \ed
 Finally we observe the following:

 \bd
 {dx \over dt} = yz= cn(t,k)\  dn(t,k)=\sqrt { 1 -x^2} \sqrt{ 1 -k^2
 x^2} \  . \ed
 Therefore
 \bd
 {dt \over dx} = { 1 \over \sqrt{ (1-x^2)(1-k^2 x^2) } } \ . \ed
 Thus we end-up with the classical definition of  elliptic
 integrals.

\section{Hamiltonian Approach}

On $\r^4$ with coordinates $(q_1, q_2, p_1, p_2)$
we define

\be \label{ham}  H={1 \over 2} (1+k^2) p_1^2 +{ 1 \over 2} p_2^2 +{ 1\over 2}
\left( p_1 q_2 - p_2 q_1 \right)^2  \ee

 with $k \in {\bf R}$.

 Hamilton's equations become

\begin{eqnarray*}
\dot{q_1}={ \p H \over \p p_1} &=& (1+k^2) p_1 + (p_1 q_2 - p_2 q_1) q_2 \\
\dot{q_2}={ \p H \over \p p_2} &=& p_2- (p_1 q_2 - p_2 q_1) q_1 \\
\dot{p_1}=-{ \p H \over \p q_1} &=& (p_1 q_2 - p_2 q_1) p_2 \\
\dot{p_2}=-{ \p H \over \p q_2} &=& - (p_1 q_2 - p_2 q_1) p_1  \  .
\end{eqnarray*}

We see right away that the function ${ 1\over 2} \left( p_1^2
+p_2^2 \right) $ is a constant of motion. Its time derivative is $p_1 \dot{p_1}+p_2 \dot{p_2}$ which is $0$. It is  independent from $H$  and therefore the system
is integrable.

We define the following mapping $\Phi$  from  ${\bf R}^4 $ to
$ {\bf R}^3$  with coordinates $(x,y,z)$:

\begin{eqnarray*} \label{map}
x&=& p_1 \\
y&=& p_2 \\
z&=& p_1 q_2 - p_2 q_1 \ .
\end{eqnarray*}

Then $ \dot x=\dot p_1 = p_2 (p_1 q_2-p_2 q_1) =yz$, similarly
$\dot y=\dot p_2 = -xz$  and $ \dot z=-k^2 x y$. In other words the Hamiltonian system defined by $H$ is transformed onto the system (\ref{e3}).  The Hamiltonian
in the new coordinates takes the form

\bd \hat H={ 1 \over 2}(1+k^2)  x^2 +{ 1 \over 2} y^2+ { 1 \over
2} z^2 \ . \ed

The second constant of motion becomes  $ {1 \over 2} \left( x^2
+y^2 \right) $. The difference of this function from the
Hamiltonian is  $ { 1 \over 2} \left( k^2 x^2 + z^2 \right)$.
Therefore, ignoring the factor ${ 1 \over 2}$ we obtain the two
functions  $F$, $G$ of  Section 2. The new Hamiltonian has the
form  ${1 \over 2} \left( F+G \right) $.

The standard symplectic bracket in  ${\bf R}^4$ is mapped onto the
Poisson bracket

\bd \{x,y \}=0, \qquad \{ x, z \}= y,  \qquad \{ y, z \}= -x \ . \ed
Let us denote the Poisson matrix of this bracket by $\pi_1$, i.e.,
\bd
\pi_1=\begin{pmatrix} 0 & 0 & y  \cr
              0 &0 &-x \cr
              -y &x& 0   \end{pmatrix} \ . \ed

In conclusion the mapping $\Phi$  is a Poisson mapping between the symplectic bracket  on $\r^4$  and the Poisson bracket $\pi_1$ on $\r^3$.
This  Poisson bracket $\pi_1$  has a Casimir $F=x^2+y^2$,  i.e. the
bracket of  $F$ with every other function is  $0$ and therefore,
we can take  $H_1=\frac{1}{2} G={ 1 \over 2}\left( k^2 x^2 +z^2 \right)$ as the
Hamiltonian.  Hamilton's equations have the form  $\dot x_i=\{x_i,
H_1 \}$. We obtain   \bd \dot x=\{ x, H_1 \} =\{x, { 1 \over 2}
\left( k^2 x^2 +z^2 \right) \} = yz,    \qquad  \dot y=\{ y, H_1 \}=-xz, \qquad \dot z=\{ z, H_1 \}=-k^2 xy  \ . \ed

In other words we
obtain again  equations  (\ref{e3}).
Define the vector $X$ by

              \begin{align}
    X &= \begin{bmatrix}
           \dot{x} \\
            \dot{y} \\
               \dot{z}
         \end{bmatrix}
  \end{align}
  and let the vector $\nabla H_1$ be the gradient vector of $H_1$, i.e.

              \begin{align}
    \nabla H_1  &= \begin{bmatrix}
           k^2 x \\
            0 \\
               z
         \end{bmatrix} \ .
              \end{align}
              Then \bd X=\pi_1 \nabla H_1 \ed
              is equivalent to equations (\ref{e3}).

The system is bi-hamiltonian.  We can take as Hamiltonian  the function $ H_2=\frac{1}{2} F= \frac{1}{2} (x^2 +y^2)$  and define the  Poisson structure  $\pi_2$
by \bd \{x,y \}= z,    \qquad \{ x, z \} =0,   \qquad \{ y, z \}=  k^2 x  \ . \ed

Then we easily verify that
\bd
X=\pi_1  \,  \nabla H_1  = \pi_2 \, \nabla  H_2 \ .
\ed

  Another way to obtain a bi-hamiltonian formulation is to choose  $ { 1 \over 2} (x^2 +y^2)$  and the following   Poisson  bracket

  \be  \label{poisson3}  \{x,y \}= z \qquad \{ x, z \} ={ 1\over 2} k^2
y \qquad \{ y, z \}= { 1 \over 2} k^2 x  \ . \ee

The type of brackets we have obtained  in three  dimensions so far are all Poisson (and hence also compatible) without any constraints.  As we will see in the following sections, in higher dimensions it is  no longer the case.

We may  generalize  system (\ref{ham}) as follows:  Consider the Hamiltonian system on $\r^6$ with coordinates $(q,p)$, equipped with the standard Poisson bracket,  which is defined be the function

\be \label{ham3}  H={1 \over 2}  p_1^2 + {1 \over 2}  p_2^2+ {1 \over 2}  p_3^2 +{ 1\over 2}
\left( p_1 q_2 - p_2 q_1 \right)^2 +{ 1\over 2}
\left( p_1 q_3 - p_3 q_1 \right)^2 +{ 1\over 2}
\left( p_2 q_3 - p_3 q_2 \right)^2  \  \ . \ee

This system is integrable.  Extending our earlier construction  we  define the following Poisson map  from $\r^6  \to \r^6$:

\begin{eqnarray} \label{map}
x_1&=& p_1  \\
x_2&=& p_2 \nonumber\\
x_3&=& p_3 \nonumber\\
y_1&=&  p_3 q_2-p_2 q_3 \nonumber\\
y_2&=& p_1 q_3 - p_3 q_1 \nonumber\\
y_3&=&  p_2 q_1 -p_1 q_2 \nonumber\ .
\end{eqnarray}

The symplectic bracket of rank 6 maps onto the Poisson bracket of rank 4 with Poisson matrix

\bd \pi=
\begin{pmatrix}
0&X\\
X&Y
\end{pmatrix}  \ \ ,
\ed

where

\bd X=
\begin{pmatrix}
0&x_3&-x_2\\
-x_3&0&x_1\\
x_2&-x_1&0
\end{pmatrix}  \ \ ,
\ed

and

\bd Y=
\begin{pmatrix}
0&y_3&-y_2\\
-y_3&0&y_1\\
y_2&-y_1&0
\end{pmatrix}  \ \ .
\ed
  The bracket $\pi$  has two Casimirs $f_1=x_1^2+x_2^2+x_3^2$ and $f_2=x_1 y_1+x_2 y_2+x_3 y_3$.  The
Poisson bracket is easily seen to be  the Lie-Poisson bracket on $e(3)=so(3) \rtimes \r^3$. The system is a
very special case of the Clebsch system where all parameters are set equal
to 1,  see \cite{pol}.

More generally, by introducing coefficients in the Hamiltonian $H$ and using the transformation (\ref{map})  we get a new system with the Clebsch Hamiltonian

\bd h=\frac{1}{2} \left( \lambda_1 x_1^2+\lambda_2 x_2^2+\lambda_3 x_3^2+ +\kappa_1 y_1^2+\kappa_2 y_2^2+\kappa_3 y_3^2 \right) \ . \ed
The mapping (\ref{map}) is  a symplectic realization of the Poisson bracket $\pi$ which  is the Lie-Poisson structure on $e(3)$   with Casimirs $f_1$, $f_2$ as before.   When the constants satisfy the condition
\bd  \frac{ \lambda_2-\lambda_3}{\kappa_1}  + \frac{ \lambda_3-\lambda_1}{\kappa_2}+\frac{ \lambda_1-\lambda_2}{\kappa_3}=0 \ed

we obtain one of the well-known integrable cases of geodesic flow on $e(3)=so(3) \rtimes \r^3$.
The equations of motion   can be written in the vector form

\bd \dot{y}=x \wedge \frac{\partial h}{\partial x}- y \wedge \frac{ \partial h }{\partial y}  \ed
\bd \dot{x}=x \wedge \frac{ \partial h }{\partial y}  \    \ed

where $x=(x_1, x_2,x_3)$,  $y=(y_1, y_2, y_3)$  and  $\wedge$ denotes the cross product on $\r^3$. The extra integral found by Clebsch is
\bd f_3=c\left(y_1^2+y_2^2+y_3^2+\kappa_1 x_1^2 +\kappa_2 x_2^2+\kappa_3 x_3^2\right)  \ , \ed
where
\bd c=\frac { \kappa_1 (\kappa_2-\kappa_3)}{\lambda_2-\lambda_3}=\frac { \kappa_2 (\kappa_3-\kappa_1)}  {\lambda_3-\lambda_1}=\frac { \kappa_3 (\kappa_1-\kappa_2)} {\lambda_1-\lambda_2} \ .  \ed
For more details on this system see \cite{pol, borisov1, borisov2, dragovic,  dubrovin, perelomov}.

\section{Pl\"ucker coordinates}

Let $U$ be a $k$-dimensional subspace of the $n$-dimensional Euclidean space $V=F^n$ where $F$ is a field of characteristic zero. The set of all such $U$ is the Grassmannian $G(k,n)$ or $G(k,V)$.    Let $N=\binom {n}{ k} -1$.  Suppose that $U$ is spanned by the columns of a $n \times k$ matrix $A$.  For every subset $I$  of $\{ 1, \dots, n\}$ of cardinality $k$, let $p_I$ be the $k\times k$ subdeterminant of $A$ defined by the rows of $I$. The vector $p=(p_I),  |I|=k$ is called the vector of Pl\"ucker coordinates  of $U$. Since at least one of the coordinates of $p$ is non-zero this defines a point in $\mathbb{P}^N$. In this paper, the most relevant  case is when $k=2$.  In the following running example we take $n=4$.

\begin{example}

  Consider the matrix $A$ given by

\bd A=
\begin{pmatrix}
x_1&y_1\\
x_2&y_2\\
x_3&y_3\\
x_4&y_4
\end{pmatrix}  \ \ .
\ed
Then the Pl\"ucker  vector $p=(p_{ij})$ has six components.  The coordinate $p_{ij}$ corresponds to the sub-determinant defined by rows $i$ and $j$.
\be p_{ij}=x_i y_j-x_j y_i, \ \ \ \  1 \le i<j \le 4   \  \ . \ee
We think of this vector as a point in $\mathbb{P}^5$.  Not every vector in $\mathbb{P}^5$ is the Pl\"ucker vector of a $2$-dimensional subspace of $F^4$.
\end{example}
 Another way to view  the Pl\"ucker imbedding is to consider the map
 \bd \gamma : G(k, n) \to \mathbb{P} \left(\bigwedge^{\raisebox{-0.4ex}{\scriptsize $k$}} V \right) \ed
 given by
 \bd Span(v_1, \dots, v_k ) \to [v_1 \wedge \dots \wedge v_k]   \ . \ed
 This map $\gamma$ is injective and the image of $G(k, n)$ is closed. An element in the image of $\gamma$ is called totally decomposable and it can be written in the form $v_1 \wedge \dots \wedge v_k$.  A simple argument shows that if $n=3$ then every non-zero element of $\bigwedge^{\raisebox{-0.4ex}{\scriptsize $2$}}V $  is totally decomposable. In dimension $n=4$ this is not the case.

\begin{example}
Suppose dim $V=4$ and take a basis $e_1, e_2, e_3, e_4$.   Then every element $v \in \bigwedge^{\raisebox{-0.4ex}{\scriptsize $2$}} V$ can be written as
\bd v=p_{12} (e_1 \wedge e_2 ) +p_{13}(e_1 \wedge e_3 )+p_{14} (e_1 \wedge e_4 )+p_{23} (e_2 \wedge e_3 )+p_{24} (e_2 \wedge e_4 )+p_{34} (e_3 \wedge e_4 )  \ .  \ed
In order for $v$ to be totally decomposable,  we should have $v \wedge v =0$, i.e.,
\bd v \wedge v =2(p_{12} p_{34}-p_{13} p_{24} +p_{14} p_{23} )=0  \ .  \ed
Therefore the image of the Pl\"ucker embedding of $G(2,4)$ into $\mathbb{P}^5$ satisfies the homogeneous quadric equation (the Klein quadric)
\be \label{plucker4}  p_{12} p_{34}-p_{13} p_{24} +p_{14} p_{23}=0  \ . \ee

\end{example}

Let $U$ be a $k$-dimensional subspace of $V$ spanned by $v_1,  \dots, v_k$ and $(e_1, e_2, \dots, e_n)$ a basis of $V$.   Then the vector of coefficients $p_{i_1 \dots i_k}$,  where $i_1 < i_2< \dots < i_k$, in the basis representation
\bd v_1 \wedge \dots \wedge v_k =\sum p_{i_1 \dots i_k} e_{i_1}  \wedge \dots e_{i_k}  \ , \ed
equals the Pl\"ucker coordinates of $U$ in homogeneous coordinates, i.e. both vectors denote the same point in $\mathbb{P}^N$.

Given a fixed $\omega \in \bigwedge^{\raisebox{-0.4ex}{\scriptsize $k$}} V$ we examine the linear map

\bd f_{\omega}: V \to \bigwedge^{\raisebox{-0.4ex}{\scriptsize $k+1$}} V   \ , \ed
given by
\bd v  \to v \wedge \omega \ . \ed

By choosing  canonical bases for $V$ and $\bigwedge^{\raisebox{-0.4ex}{\scriptsize $k+1$}} V$ in lexicographic order, we obtain the corresponding matrix of $f_{\omega}$ which we denote by
$A_{\omega}$.  Then one can prove:
\begin{proposition}
The following are equivalent

\noindent
(i) $ \omega$ is totally decomposable

\noindent
(ii)dim  Ker$f_{\omega}=k$

\noindent
(iii) rank  $A_{\omega}=n-k$.

\end{proposition}

\begin{example}
We look again at the case $G(2,4)$ for an illustration. A vector $\omega \in \bigwedge^{\raisebox{-0.4ex}{\scriptsize $2$}} V $ can be written in the form
\bd \omega =p_{12} (e_1 \wedge e_2 ) +p_{13}(e_1 \wedge e_3 )+p_{14} (e_1 \wedge e_4 )+p_{23} (e_2 \wedge e_3 )+p_{24} (e_2 \wedge e_4 )+p_{34} (e_3 \wedge e_4 )  \ .  \ed
 Using the  basis $e_1 \wedge e_2 \wedge e_3, e_1 \wedge e_2 \wedge e_4, e_1 \wedge e_3 \wedge e_4, e_2 \wedge e_3 \wedge e_4$  of     $\bigwedge^{\raisebox{-0.4ex}{\scriptsize $3 $}} V $ the representation matrix
$A_{\omega}$    is

\bd A_{\omega}=
\begin{pmatrix}
p_{23}&-p_{13}&p_{12} &0 \\
p_{24}&-p_{14}&0 &p_{12} \\
p_{34}&0&-p_{14} &p_{13} \\
0&p_{34}&-p_{24} &p_{23}
\end{pmatrix}  \ \ .
\ed
The vector $\omega$ defines a Pl\"ucker vector of a line in $\mathbb{P}^3$ if and only if this matrix $A_{\omega}$  has rank 2. That means that all $3 \times 3$ minors of this matrix should vanish.  The image of $G(2,4)$ under the Pl\"ucker embedding is the common zero locus of the sixteen $3 \times 3$ minors of $A_{\omega}$. For example

\bd
\begin{vmatrix}
p_{23}&-p_{13}&p_{12}  \\
p_{24}&-p_{14}&0  \\
p_{34}&0&-p_{14}
\end{vmatrix}  =p_{14} (p_{14} p_{23}  -p_{13} p_{24} +p_{12} p_{34})=0  \  \ .
\ed
Indeed, four of the minors are zero,  and the rest are all multiples of the quadratic  polynomial $p_{12} p_{34}-p_{13} p_{24} +p_{14} p_{23}$ which  we found earlier.

\end{example}

More generally the Pl\"ucker relations for $G(2, n)$  are given by the vanishing of quadratic polynomials of  the form

\be \label{diagonal}  p_{ij} p_{kl} -p_{ik} p_{jl}+ p_{jk} p_{il} =0  \ \   \ee
where $1 \le i<j<k<l \le n$.

For later use we note that a line $l$ intersects a line $l^{\prime}$ in $\mathbb{P}^3$ if their Pl\" ucker coordinates $p$ and  $p^{\prime}$ satisfy

\be \label{line}   p_{12} p_{34}^{\prime} -p_{13} p_{24}^{\prime} + p_{14} p_{23}^{\prime} +p_{23} p_{14}^{\prime}-p_{24} p_{13}^{\prime} +p_{34} p_{12}^{\prime} =0  \  . \ee
See \cite{joswig} for details.

\section{Generalization to dimension 4}

We generalize the Poisson bracket (\ref{poisson3}) from three to four dimensions as follows:

\begin{eqnarray} \label{poisson4}
\{x_1, x_2 \} &=& \pi_{12} x_3 x_4 \\ \nonumber
\{x_1, x_3 \}&=& \pi_{13}  x_2 x_4 \\  \nonumber
\{x_1, x_4 \} &=& \pi_{14} x_2 x_3   \\  \nonumber
\{x_2, x_3 \} &=& \pi_{23} x_1  x_4  \\  \nonumber
\{x_2, x_4 \} &=& \pi_{24} x_1 x_3  \\  \nonumber
\{x_3, x_4 \} &=& \pi_{34} x_1 x_2   \label{e5} \ .
\end{eqnarray}

A simple calculation shows that the Jacobi identity is satisfied if and only if  the following condition holds

\be \pi_{12} \pi_{34}-\pi_{13} \pi_{24} +\pi_{14} \pi_{23} =0  \ .  \label{e6}  \ee
We recognize this equation as the Pl\"ucker relation (\ref{plucker4}).

\begin{rem}
This type of bracket appears in \cite[p. 23]{pol2} as a generalization of the Sklyanin bracket, and where it is noted that it gives a 5-dimensional family of quadratic Poisson structures of rank two.

\end{rem}

\begin{example} \label{ex3}

Define
\bd \pi_{12}=-1,  \ \ \pi_{13}=2,  \ \ \pi_{14}=0,   \pi_{23}=-2,   \ \  \pi_{24}=-k^2  \ \  \pi_{34}=2 k^2    \  \ .  \ed
Take $H=\frac{1}{2} (x_1^2 +x_2^2 +x_3^2)$.   Then the equations of motion become
\begin{eqnarray}
\dot x_1 &=&   x_2 x_3 x_4 \\ \nonumber
 \dot x_2&=&  -  x_1 x_3 x_4 \\  \nonumber
 \dot x_3&=& 0   \\  \nonumber
 \dot x_4&=& -k^2 x_1 x_2 x_3
   \label{e4} \ .
\end{eqnarray}
The functions  $f=k^2 x_1^2+x_4^2$ and $g=2x_1^2 +2x_2^2 +  x_3^2 $  are Casimirs of the Poisson bracket.
 We remark  that the Casimirs may be determined  by finding the Kernel of the constant matrix $(\pi_{ij})$ which consists of the vectors $(k^2, 0,0, 1)$, $(2,2,1,0)$.
Also, note  that the choice $x_3=1$, $x_1=x$, $x_2=y$ and $x_4=z$ gives equations (\ref{e3}).

\end{example}

\begin{example}{The Sklyanin bracket \cite{sklyanin}} \label{ex4}

Let $j_1, j_2, j_3$ be constants. The choice
\bd \pi_{12}=j_2-j_3,  \ \ \pi_{13}=j_3-j_1,  \ \ \pi_{14}=j_1-j_2,   \pi_{23}=1,  \pi_{34}=-1 \ \  \pi_{24}=1  \  \ ,  \ed
leads to the Sklyanin bracket. The Casimirs  are   $x_1^2+j_1 x_2^2+j_2 x_3^2+j_3 x_4^2$ and $x_2^2+x_3^2+x_4^2$.

\end{example}

\bigskip

The Poisson bracket in the two examples can be obtained in a different way  using the Jacobian type Poisson formula (\ref{det}).
In the case $n=4$ the formula (\ref{det}) takes the form

\bd \{ F,  G \} ={\rm det} (\nabla F,  \nabla G, \nabla f, \nabla g )  \ , \ed
where $f$ and $g$ are two arbitrary independent  functions on $\r^4$.

Let us take  $f=\frac{1}{2} (\alpha_1  x_1^2 + \alpha_2 x_2^2 +\alpha_3 x_3^2 +\alpha_4 x_4^2 )$  and
$g=\frac{1}{2} (\beta_1  x_1^2 + \beta_2 x_2^2 +\beta_3 x_3^2 +\beta_4 x_4^2 )$.
Consider the Jacobian bracket generated by $f$ and $g$.
For example,
\bd \{x_1,  x_2 \}={\rm det}
\begin{pmatrix}
1&0&\alpha_1 x_1&\beta_1 x_1\\
0&1&\alpha_2 x_2&\beta_2 x_2\\
0&0&\alpha_3 x_3&\beta_3 x_3\\
0&0&\alpha_4 x_4&\beta_4 x_4\\
\end{pmatrix}  =(\alpha_3 \beta_4-\alpha_4 \beta_3)x_3 x_4 \  .
\ed

\begin{proposition} \label{fg}
The Jacobian Poisson bracket on $\r^4$ generated by the two Casimirs $f$, $g$  is of the form (\ref{poisson4}).  Conversely, the Poisson bracket (\ref{poisson4}) is a Jacobian type bracket generated by two Casimirs.

\end{proposition}

\proof

Using (\ref{det}) we obtain
\begin{eqnarray*} \label{plucker}
\pi_{12} &=&   (\alpha_3 \beta_4-\alpha_4 \beta_3)=\pi_{34}^{\prime} \\ \nonumber
 \pi_{13} &=&   (\alpha_4 \beta_2-\alpha_2 \beta_4)=-\pi_{24}^{\prime} \\ \nonumber
 \pi_{14} &=&   (\alpha_2 \beta_3-\alpha_3 \beta_2)=\pi_{23}^{\prime} \\ \nonumber
 \pi_{23} &=&   (\alpha_1 \beta_4-\alpha_4 \beta_1)=\pi_{14}^{\prime} \\ \nonumber
 \pi_{24} &=&   (\alpha_3 \beta_1-\alpha_1 \beta_3)=-\pi_{13}^{\prime} \\ \nonumber
 \pi_{34} &=&   (\alpha_1 \beta_2-\alpha_2 \beta_1)=\pi_{12}^{\prime}  \  \ .
\end{eqnarray*}
Then we have
\bd \pi_{12} \pi_{34}-\pi_{13} \pi_{24} +\pi_{14} \pi_{23}=\pi_{34}^{\prime}\pi_{12}^{\prime}-\pi_{24}^{\prime}\pi_{13}^{\prime}+\pi_{23}^{\prime}\pi_{14}^{\prime}=0  \ \ . \ed

Therefore the Jacobi identity is satisfied. We obtain a Jacobian Poisson bracket of rank 2 and of the form  (\ref{poisson4}).

Conversely, if

\bd  \pi=
\begin{pmatrix}
0& \pi_{12} x_3 x_4&\pi_{13} x_2 x_4 &\pi_{14} x_2 x_3 \\
-\pi_{12} x_3 x_4&0&\pi_{23} x_1 x_4 &\pi_{24} x_1 x_3 \\
-\pi_{13} x_2 x_4&- \pi_{23} x_1 x_4&0 &\pi_{34} x_1 x_2 \\
-\pi_{14} x_2 x_3& -\pi_{24} x_1 x_3&-\pi_{34} x_1 x_2 &0
\end{pmatrix}  \ \ ,
\ed
then det $\pi$ is equal to $x_1^2 x_2^2 x_3^2 x_4^2 (\pi_{12} \pi_{34}-\pi_{13} \pi_{24} +\pi_{14} \pi_{23})^2$.  The  bracket has rank 2 if and only if
$\pi_{12} \pi_{34}-\pi_{13} \pi_{24} +\pi_{14} \pi_{23}=0$. Therefore it is Poisson if and only if it has rank 2. It follows from the fact that the Pl\" ucker map is injective that
we can define   $f=\frac{1}{2} (\alpha_1  x_1^2 + \alpha_2 x_2^2 +\alpha_3 x_3^2 +\alpha_4 x_4^2 )$  and $g=\frac{1}{2} (\beta_1  x_1^2 + \beta_2 x_2^2 +\beta_3 x_3^2 +\beta_4 x_4^2 )$ so that the Poisson bracket $\pi$ is the Jacobian Poisson bracket generated by $f$ and $g$.  We prove this explicitly in the case $\pi_{23} \not=0$.   We take $\alpha_1=0$ and $\beta_4=0$.  Then we have

\begin{eqnarray*}
\pi_{12} &=&   -\alpha_4 \beta_3 \\ \nonumber
 \pi_{13} &=&   \alpha_4 \beta_2 \\ \nonumber
 \pi_{14} &=&   \alpha_2 \beta_3-\alpha_3 \beta_2 \\ \nonumber
 \pi_{23} &=&   -\alpha_4 \beta_1 \\ \nonumber
 \pi_{24} &=&   \alpha_3 \beta_1 \\ \nonumber
 \pi_{34} &=&   -\alpha_2 \beta_1 \  \ .
\end{eqnarray*}

The fourth and fifth equation imply that $\alpha_3 \pi_{23}=-\alpha_4 \pi_{24}$.  We choose $\alpha_3=-\pi_{24}$ and $\alpha_4=\pi_{23}$.  The first equation gives $\beta_3=-\frac{ \pi_{12}}{\pi_{23}} $ and the second gives
$\beta_2=\frac{ \pi_{13}}{\pi_{23}}$. The fourth equation implies that $\beta_1=-1$ and the last equation $\alpha_2=\pi_{34}$. The consistency of the third equation follows from the Pl\"ucker relation. Therefore we may choose
 \bd f=\pi_{34} x_2^2-\pi_{24}x_3^2 +\pi_{23}x_4^2  \qquad g=\pi_{23}x_1^2-\pi_{13}x_2^2 +\pi_{12}x_3^2 \ . \ed                This completes the proof.

\qed

Given a system of the form
\begin{eqnarray} \label{system4}
\dot{x}_1 &=&  c_1 x_2 x_3 x_4 \\ \nonumber
\dot{x}_2 &=&  c_2 x_1 x_3 x_4 \\ \nonumber
\dot{x}_3 &=&  c_3 x_1 x_2 x_4 \\ \nonumber
 \dot{x}_4 &=&  c_4 x_1 x_2 x_3  \  \
\end{eqnarray}
where the $c_i$ are constants  all  different than zero, it has the obvious integrals $f_1=c_2 x_1^2-c_1 x_2^2$ and $f_2=c_4 x_3^2-c_3 x_4^2$.  Using the Jacobian formula (generated by $f_1$ and $f_2$)  we obtain a Poisson bracket of the form (\ref{poisson4}) where

\bd \pi_{12}=0,  \ \ \pi_{13}=-4c_1 c_3, \ \ \pi_{14}=-4c_1 c_4,   \pi_{23}=-4c_2 c_3,   \ \  \pi_{24}=-4c_2 c_4, \ \  \pi_{34}=0    \  \ .  \ed
Using this Poisson bracket and taking as
Hamiltonian the function
\bd H=\frac{1}{16} \left( \frac{x_1^2}{c_1}+\frac{x_2^2}{c_2}-\frac{x_3^2}{c_3}-\frac{x_4^2}{c_4} \right) \ed
we obtain equations (\ref{system4}). Therefore the system (\ref{system4}) is always a completely integrable Hamiltonian system.  This procedure, of course, generalizes easily to  higher dimensions.

Recall the definition of compatible Poisson brackets. Let  $ \pi_1$, $ \pi_2$ be two Poisson structures on $M$.
If  $\pi_1 +\pi_2$ is also Poisson then the two tensors   form a {\it Poisson pair}  on $M$. The
corresponding Poisson brackets are called {\it compatible}. We would like to find a condition so that two brackets of the form (\ref{poisson4}) are compatible.
Consider two brackets in $\r^{4}$ of the form  (\ref{poisson4})  defined by functions
$\pi_{ij}$ and  $\pi_{ij}^{\prime}$.
When are they compatible?   A simple computation of the Jacobi identity  shows that the following identity should be satisfied:
\bd  \pi_{12} \pi_{34}^{\prime} -\pi_{13} \pi_{24}^{\prime} + \pi_{14} \pi_{23}^{\prime} +\pi_{23} \pi_{14}^{\prime}-\pi_{24} \pi_{13}^{\prime} +\pi_{34} \pi_{12}^{\prime} =0  \  . \ed
Comparing this equation with equation (\ref{line}) we immediately have the following:

\begin{proposition} \label{comp}
Consider two Poisson brackets $\pi_{ij} $ and $\pi_{ij}^{\prime}$ associated to two lines $l$ and  $l^{\prime}$ in $\mathbb{P}^3$.  Then   $\pi_{ij} $ and $\pi_{ij}^{\prime}$ are compatible if and only if the line $l$ intersects the  line $l^{\prime}$ in $\mathbb{P}^3$

\end{proposition}

This Proposition generalizes in a straightforward manner in higher dimensions.  The Poisson brackets of examples (\ref{ex3}) and (\ref{ex4}) are not compatible.

\section{Generalization to $\r^n$}

More generally we  consider on $\r^n$  a bracket of the form

\be \label{poissonn}  \{ x_i,  x_j \} = \pi_{ij} x_1 x_2 \cdots \hat{x_i} \cdots \hat{x_j}  \cdots x_n   \ .  \ee

 Let us denote this Poisson tensor by $p=(p_{ij})$.
As in dimension $4$ the Jacobi identity is satisfied if and only if the  Pl\"ucker relations hold.

\begin{theorem}
The bracket $(p_{ij})$   is Poisson if and only if the $\pi_{ij}$ satisfy the Pl\"ucker relations (\ref{diagonal}).
\end{theorem}

\proof
We may assume $i<j<k$.
It suffices to  check the Jacobi identity for coordinate functions $x_i, x_j, x_k$.

\begin{gather*}
 \{x_i, \{x_j, x_k \} \}+\{ x_j, \{x_k, x_i \} \} + \{x_k, \{x_i,  x_j \} \}  =   \\
 \{ x_i,  \pi_{jk} x_1 x_2 \cdots \hat{x_j} \cdots \hat{x_k} \cdots  x_n \} -\{x_j,  \pi_{ik} x_1 x_2 \cdots \hat{x_i} \cdots \hat{x_k} \cdots  x_n \} + \\
  \{x_k,  \pi_{ij} x_1 x_2 \cdots \hat{x_i} \cdots \hat{x_j} \cdots  x_n \}  =  \\
 =   \pi_{jk} \left(  \sum_{l\not = j, k} \{x_i, x_l\} x_1 x_2  \cdots  \hat{x_l}  \cdots \hat{x_j} \cdots \hat{x_k} \cdots  x_n  \right)    \\
  -\pi_{ik} \left(  \sum_{l\not = i, k} \{x_j, x_l\} x_1 x_2  \cdots  \hat{x_l}  \cdots \hat{x_i} \cdots \hat{x_k} \cdots  x_n  \right) +    \\
  \pi_{ij} \left(  \sum_{l\not = i, j} \{x_k, x_l\} x_1 x_2  \cdots  \hat{x_l}  \cdots \hat{x_i} \cdots \hat{x_j} \cdots  x_n  \right) \ .
\end{gather*}
We examine this expression for fixed $l$.   It is equal to

\begin{gather*}
   \pi_{jk} \left(  {\pi_{il}}   x_1 x_2  \cdots  \hat{x_i}  \cdots \hat{x_l}  \cdots  x_n  \right)_{l\not = j, k}  x_1 x_2  \cdots  \hat{x_l}  \cdots \hat{x_j} \cdots \hat{x_k} \cdots  x_n -  \\
 -\pi_{ik} \left(  {\pi_{jl}}  x_1 x_2  \cdots  \hat{x_j}  \cdots \hat{x_l}  \cdots  x_n  \right)_{l\not = i, k} x_1 x_2  \cdots  \hat{x_l}  \cdots \hat{x_i} \cdots \hat{x_k} \cdots  x_n +  \\
 +  \pi_{ij} \left(  {\pi_{kl}}  x_1 x_2  \cdots  \hat{x_k}  \cdots \hat{x_l}  \cdots  x_n  \right)_{l\not = i, j} x_1 x_2  \cdots  \hat{x_l}  \cdots \hat{x_i} \cdots \hat{x_j} \cdots  x_n   \\
 =(\pi_{jk} \pi_{il} -\pi_{ik} \pi_{jl}+\pi_{ij} \pi_{kl} )  x_1^2  x_2^2 \cdots \hat{x_l} \cdots x_n^2  x_i x_j x_k  \   .
\end{gather*}
Since the expression involves homogeneous polynomials,  the Jacobi identity is satisfied only if  the coefficients all vanish.
Therefore we get
\bd \pi_{jk} \pi_{il} -\pi_{ik} \pi_{jl}+\pi_{ij} \pi_{kl} =0  \ .  \ed

\ \ \
\qed

The vector defined by  $\pi_{ij}$ represents a point in $\mathbb{P}^N$  where $N=\frac{(n-2)(n+1)}{2}$. It is the image of a line in $\mathbb{P}^{n-1}$ under the Pl\" ucker imbedding.

We now show  that the rank of the bracket (\ref{poissonn}) is  2.

\begin{theorem}
The rank of the bracket $(p_{ij})$  is at most 2.
\end{theorem}

\proof
 Suppose that the $\pi_{ij}$ satisfy the Pl\" ucker relations.  Then we may write $\pi_{ij}=\alpha_{i} \beta_{j}-\alpha_{j} \beta_{i}$.
Define the vector fields
\bd X=\sum_i \frac{\alpha_i}{x_i} \frac{\partial } {\partial x_i}  \qquad Y=\sum_j \frac{\beta_j}{x_j} \frac{\partial } {\partial x_j} \  . \ed

We note that the two vector fields  commute, i.e.,  $[X, Y]=0$. It follows that the tensor $X \wedge Y$ is  a Poisson tensor of rank 2. Let
 \bd \rho=X \wedge Y=\left( \frac{\pi_{ij}}{ x_i x_j} \right) \ .  \ed

  Since  the rank is 2 we may multiply $\rho$  by an arbitrary function without altering the Jacobi identity.  Let
$\Psi=x_1 x_2 \dots x_n$.  We verify that $\Psi\rho=\Psi X \wedge Y$ is precisely the Poisson tensor $p=(p_{ij})$.

\qed

\begin{rem}
A completely different proof of this result can be found in \cite{benito} 
\end{rem}

If we   have a Poisson bracket on $\r^n$  and we define two matrices $A$ and $B$ by
$A_{ij}:=\left(\{x_i,x_j\}\right)$  and $B_{ij}:=\left(\{x_i^2,x_j^2\}\right)$ then at a generic point (i.e.
where all $x_i$ are different from zero) both matrices have the same rank.    Let us use  for the first matrix the bracket $\rho_{ij} =\left( \frac{\pi_{ij}}{ x_i x_j} \right)$.  Then the second matrix
satisfies $\{x_i^2,x_j^2\}=4\pi_{ij}$. In conclusion:  the constant Poisson matrix $\pi_{ij}$ and the Poisson matrix $p_{ij}$  both have the same rank, i.e.,  2.

We now examine the problem of determining the Casimirs of the bracket (\ref{poissonn}). The  constant matrix $\pi_{ij}$ will be used in the computation of the Casimirs.

Let $\alpha $ be a vector in the Kernel of the constant matrix $\pi_{ij}$, i.e.
\bd (\pi_{ij}) \cdot \alpha =0  \ . \ed
Equivalently
\bd \sum_{k} \pi_{ik} \alpha_k =0 \ . \ed
Define the quadratic function $f$ by the formula

\bd f=\sum_{k}\alpha_k x_k^2  \ . \ed

\begin{proposition}
The function $f$ is a Casimir of the Pl\" ucker bracket $(p_{ij})$.
\end{proposition}

\proof
It is enough to show that it is a Casimir for the rational  bracket
\bd \rho_{ij} =\left( \frac{\pi_{ij}}{ x_i x_j} \right) \ . \ed
We compute:
\bd (\rho_{ij}) \nabla f= \sum_k \frac{ \pi_{ij} } { x_i x_k} \cdot 2 \alpha_k x_k = \frac{2}{x_i} \left( \sum_{k} \pi_{ik} \alpha_k \right) =0 \ .   \ed

\ \ \
\qed

\begin{example}
Let us examine  in detail the case $n=5$.  In this case there are five Pl\" ucker relations:

\begin{eqnarray*}
\pi_{12} \pi_{34}-\pi_{13} \pi_{24} +\pi_{14} \pi_{23} &=& 0 \\ \nonumber
\pi_{12} \pi_{35}-\pi_{13} \pi_{25} +\pi_{15} \pi_{23} &=& 0 \\ \nonumber
 \pi_{12} \pi_{45}-\pi_{14} \pi_{25} +\pi_{15} \pi_{24} &=& 0 \\ \nonumber
 \pi_{13} \pi_{45}-\pi_{14} \pi_{35} +\pi_{15} \pi_{34} &=& 0 \\ \nonumber
 \pi_{23} \pi_{45}-\pi_{24} \pi_{35} +\pi_{25} \pi_{34} &=& 0  \ .
\end{eqnarray*}
To determine  the Casimirs we need to find the Kernel of the matrix

\bd  \pi=
\begin{pmatrix}
0& \pi_{12} &\pi_{13}  &\pi_{14} &\pi_{15} \\
-\pi_{12} &0&\pi_{23} &\pi_{24}& \pi_{25}  \\
-\pi_{13} &- \pi_{23} &0 &\pi_{34}&\pi_{35}  \\
-\pi_{14} & -\pi_{24}&-\pi_{34}  &0&\pi_{45}  \\
-\pi_{15} & -\pi_{25} & -\pi_{35}&-\pi_{45} &0
\end{pmatrix}  \ \  .
\ed
The dimension of the Kernel is three.   We may determine three  vectors in the Kernel as follows. Since the Pl\" ucker relations involve three terms we may choose a column vector with two zeros. For example, if we choose the first and last entry to be 0, then the other three  entries are simply determined by one of  the 5 Pl\" ucker relations above.  We end-up with the vector $(0, \pi_{34}, -\pi_{24}, \pi_{23}, 0)$.  Therefore, one of the Casimirs is the function
$f_{234}= \pi_{34} x_2^2 -\pi_{24} x_3^2 +\pi_{23} x_4^2  $.  In a similar fashion we get the other two Casimirs $f_{345}=\pi_{45} x_3^2 -\pi_{35} x_4^2 +\pi_{34} x_5^2 $ and $f_{123}=\pi_{23} x_1^2 -\pi_{13} x_2^2 +\pi_{12} x_3^2$.
\end{example}
This example suggests the following result:

\begin{theorem}
The functions
$f_{ijk}=\pi_{jk} x_i^2 -\pi_{ik} x_j^2 +\pi_{ij} x_k^2$ are Casimirs of the  Poisson bracket (\ref{poissonn}).
\end{theorem}

\proof

A simple calculation gives
\bd
	\{f_{ijk},  x_l \} =
	\left\{
	\begin{array}{rl}
		  0	& \textrm{if } l \in \{ i, j,k \}\\
				 2x_1 x_2 \cdots \hat{x_l} \cdots   \cdots x_n(\pi_{il} \pi_{jk}-\pi_{jl} \pi_{ik} +\pi_{ij} \pi_{kl}) 	& \textrm{if }  l \notin \{i,j,k \}
	\end{array}
	\right.
\ed
The result follows  from the Pl\" ucker relations which we assume to hold. Note that in the formula if $l <i$  we replace $\pi_{il}$ by $-\pi_{li}$.

\qed

\begin{example} {Double-elliptic system.} (see \cite{rubtsov, odesskii}. )

The Hamiltonian of the two-particle double-elliptic system in the form of \cite{braden}  is
\bd H(p,q)=\alpha(q|k)cn\left( p \beta(q|k)|\frac{\tilde{k} \alpha(q|k)}{\beta(q|k)} \right) \ed
where $\alpha(q|k) =\sqrt{ 1 + \frac{ g^2}{sn^2 (q|k)}}$ and $\beta(q|k)=\sqrt{ 1 + \frac{ g^2 \tilde{k} }{sn^2 (q|k)}}$  coincides with $x_5$.

We choose the following system of four quadrics in $\r^6$

\begin{eqnarray*}
x_1^2-x_2^2 &=&  1 \\ \nonumber
 x_1^2-x_3^2 &=&  k^2 \\ \nonumber
 -g^2 x_1^2+x_4^2 -x_5^2 &=&  1 \\ \nonumber
 -g^2 x_1^2+x_4^2+\tilde{k}^{-2} x_6^2&=&  \tilde{k}^{-2} \  \ .
\end{eqnarray*}
Using the bracket (\ref{det}) we obtain a Poisson bracket of the form (\ref{poissonn}). Up to a factor of $\frac{16}{k^2}$  some non-trivial brackets are
\begin{eqnarray*}
\{x_1, x_5 \} &=&   -x_2 x_3 x_4 x_6 \\ \nonumber
 \{ x_2, x_5 \}&=&   -x_1 x_3 x_4 x_6 \\  \nonumber
\{x_3, x_5 \} &=& -x_1x_2 x_4 x_6   \\  \nonumber
\{x_4, x_5 \}&=&-g^2 x_1 x_2 x_3 x_6
   \label{e4} \ .
\end{eqnarray*}
The variable $x_5$ plays the role of the Hamiltonian. The fact that $\{x_5, x_6 \}=0$ shows that $x_6$ is a constant of motion.
Setting the integrals $x_5$ and $x_6$ to constants we obtain the following system of equations in $\r^4$:

\begin{eqnarray*}
\dot x_1 &=&   x_2 x_3 x_4 \\ \nonumber
 \dot x_2&=&    x_1 x_3 x_4 \\  \nonumber
 \dot x_3&=& x_1 x_2 x_4   \\  \nonumber
 \dot x_4&=& g^2 x_1 x_2 x_3
   \label{e4} \ .
\end{eqnarray*}
This system has the form of Fairlie elegant integrable  system  \cite{fairlie}. Of course the integrals survive after reduction, i.e. $x_1^2-x_2^2$,  $x_1^2-x_3^2$, $x_4^2-g^2 x_1^2$ are first integrals of the reduced system.
\end{example}

\begin{example}
Define a bracket $p_{ij}$  of the form (\ref{poissonn}) with  $n=6$ by choosing
\bd \pi_{12}=\pi_{13}=\pi_{14}=\pi_{16} =\pi_{23}=\pi_{24}=\pi_{25}=\pi_{26}=\pi_{35}=\pi_{45} =1 \ ,  \ed
\bd \pi_{15}=\pi_{34}=\pi_{36}=\pi_{46}=0 \ , \ed
\bd \pi_{56}=-1  \  . \ed
The bracket has the following four Casimirs:

\begin{eqnarray*}
f_1 & =&\frac{1}{2} (x_1^2-x_2^2+x_3^2) \\
f_2 & =&\frac{1}{2} (x_4^2-x_3^2) \\
f_3 & =&\frac{1}{2} (x_6^2-x_4^2) \\
f_4 & =&\frac{1}{2} (x_6^2-x_5^2-x_2^2)  \ .
\end{eqnarray*}
Consider the Jacobian Poisson bracket generated by the four Casimirs $f_1, f_2, f_3, f_4 $.  Then one verifies that
\bd \pb{x_i,x_j}_{det}=p_{ij}  \ . \ed

\end{example}

We would like to find a condition so that two brackets of the form (\ref{poissonn}) are compatible. Proposition \ref{comp}  generalizes easily.
Consider two brackets in $\r^{n}$ of the form  (\ref{poissonn})  defined by functions
$\pi_{ij}$ and  $\pi_{ij}^{\prime}$.
In addition to the Pl\" ucker relations satisfied by $\pi_{ij}$ and  $\pi_{ij}^{\prime}$  one needs the  following identity which is easily checked:
\bd  \pi_{ij} \pi_{kl}^{\prime} -\pi_{ik} \pi_{jl}^{\prime} + \pi_{il} \pi_{jk}^{\prime} +\pi_{jk} \pi_{il}^{\prime}-\pi_{jl} \pi_{ik}^{\prime} +\pi_{kl} \pi_{ij}^{\prime} =0  \  , \ed
where $1 \le i<j<k<l \le n$.

\section*{ACKNOWLEDGMENTS}
The author would like to thank Pol Vanhaecke  for providing the proof for  the rank and for numerous   useful suggestions. The author wishes to thank Charalampos Evripidou and the anonymous reviewers of this article for their helpful comments, which led to a considerable improvement of the presentation.

\end{document}